\documentclass[conference]{IEEEtran}
\IEEEoverridecommandlockouts
\usepackage{cite}
\usepackage{amsmath,amssymb,amsfonts}
\usepackage{algorithmic}
\usepackage{graphicx}
\usepackage{textcomp}
\usepackage{xcolor}
\def\BibTeX{{\rm B\kern-.05em{\sc i\kern-.025em b}\kern-.08em
    T\kern-.1667em\lower.7ex\hbox{E}\kern-.125emX}}

\begin{document}

\title{Unveiling Ethereum's Hidden Centralization Incentives: Does Connectivity Impact Performance?}

\author{
    \IEEEauthorblockN{Mikel Cortes-Goicoechea}
    \IEEEauthorblockA{
    \small{Barcelona Supercomputing Center}\\
    Barcelona, Spain \\
    mikel.cortes@bsc.es}
    \and
    \IEEEauthorblockN{Tarun Mohandas-Daryanani}
    \IEEEauthorblockA{
    \small{Barcelona Supercomputing Center}\\
    Barcelona, Spain \\
    tarun.mohandas@bsc.es}
    \and
    \IEEEauthorblockN{Jose Luis Muñoz-Tapia}
    \IEEEauthorblockA{
    \small{U. Politécnica de Catalunya}\\
    Barcelona, Spain \\
    jose.luis.munoz@upc.edu}
    \and
    \IEEEauthorblockN{Leonardo Bautista-Gomez}
    \IEEEauthorblockA{
    \small{Status.im}\\
    Barcelona, Spain \\
    leo@status.im}
}

\maketitle

\begin{abstract}
Modern public blockchains like Ethereum rely on p2p networks to run distributed and censorship-resistant applications. With its wide adoption, it operates as a highly critical public ledger. On its transition to become more scalable and sustainable, shifting to PoS without sacrificing the security and resilience of PoW, Ethereum offers a range of consensus clients to participate in the network. In this paper, we present a methodology to measure the performance of the consensus clients based on the latency to receive messages from the p2p network. The paper includes a study that identifies the incentives and limitations that the network experiences, presenting insights about the latency impact derived from running the software in different locations.   
\end{abstract}


\section{Introduction}
\label{sec:introduction}

Ethereum\cite{eth-whitepaper} has been an important achievement on the road to ubiquitous blockchain technology. It has shown remarkable adaptability over time, leading the technical research vanguard of the blockchain industry after being the first decentralized platform that offered a general-purpose virtual machine capable of processing the so-called \emph{smart contracts}\cite{wohrer2018smart}.
Over the last five years, Ethereum has been transitioning from an energy-hungry Proof of Work (PoW)\cite{de2018bitcoin} to a more efficient and scalable Proof of Stake (PoS)\cite{saleh2021blockchain} protocol. A transition that relies on GasperFFG \cite{buterin2017casper} and RANDAO\cite{randao-code} to replace the consensus and randomness provided by PoW. 

Since \emph{the merge}\cite{cassez2022formal}, running a validator in Ethereum's ecosystem requires two codependent software: an execution layer (EL) client and a consensus layer (CL) client or beacon node. The EL client is responsible for receiving and validating transactions or smart contracts \cite{wohrer2018smart} in the execution layer, tracking the interaction between users and the Ethereum Virtual Machine (EVM) and rewarding the proposer validator with the referenced tips of each transaction.
On the other hand, the CL client is in charge of operating, validating, and recording the interaction between validators to find consensus over the chain's state. Ethereum validators run on top of these nodes, which can interact with the rest of the beacon chain network, earning rewards based on the quality of their contribution towards consensus. 

The successful transition from PoW to PoS of Ethereum\cite{kapengut2023event} implies a radical change in the consensus mechanism. Validators must actively participate in the consensus to keep finalizing previous epochs, assigning them periodical duties they must accomplish over their lifetime \cite{cortes2023autopsy}. 
However, to perceive the maximum remuneration that PoS can grant to honest participating validators, the quality of their implication has a significant weight on their reward. Each validator has the following duties to fulfill: i) attest on a slot every epoch (defined randomly on the state transition between epochs), ii) sign sync-committee duties if the validator belongs to a sync-committee, and iii) generate and propose blocks when they are chosen to do so. The validator is now in charge of proposing blocks when they become block-proposers. 

All these duties and their implication in the consensus are defined in the Ethereum specification\cite{eth-cl-specs}. However, although there is one single specification, Ethereum relies on a wide variety of implementations to introduce the resilience of the protocol. Five main clients are consolidated to participate in the network: Lighthouse, Lodestar, Nimbus, Prysm, and Teku. Each uses a different programming language to implement the networking and consensus specs. 
Even though this multiple-spec implementation significantly impacts the feature development time (extra complexity making the implementations inter-operable between clients), it makes the protocol fault-tolerant. Ensuring that with proper client diversity in the network, a bug on a single client won't break the aggregation of new blocks to the chain. However, although all the implementations are spec-compliant, the wide variety of conditions in the protocol makes some algorithms more optimal than others under specific circumstances, i.e., certain network instabilities or the resulting latency delay derived from the geo-location of the node. Of course, having a faster or wiser algorithm might get mirrored in higher rewards. Thus, in this paper, we analyze the performance differences between geographical regions from a CL client perspective, with the final intention of spotting any existing relevant performance difference that could compromise the client diversity in the network. 

From the premise that a better accomplishment of duties generally means more reward for validators, this paper analyzes the direct implications of the networking conditions on the stability and performance of the five main CL clients. We present a study that measures the duties completion of multiple clients across multiple locations in live networks, which has not been previously done to the best of our knowledge.
The contribution of this work is to identify any missed performance between different geographical locations that could compromise the network's decentralization while proving that client diversity in the network is mature enough to be achieved in production without significant sacrifices. 
We demonstrate that all locations perform similarly under standard networking and hardware conditions, achieving an average extracted reward of $80.18$\%. However, when we deploy Ethereum clients in virtual machines or less globally connected regions, the reward can drop to $74.34$\%.   

The paper is organized as follows: Section \ref{sec:related-work} introduces the state of the art of the paper, going through previously done work on the topic, Section \ref{sec:methodology} introduces all the methodology and tools used to perform the study,  Section~\ref{sec:evaluation} presents the results obtained by our study, Section \ref{sec:discussion} discusses the insights presented in the paper, and Section \ref{sec:conclusion} summarizes all the highlights of the study.

\section{Related Work}
\label{sec:related-work}

Distributed ledgers are an interesting phenomenon in the internet space. In such a critical environment where users trust the network to track their economic balances, participants must agree on a set of rules to reach a consensus over the interaction with the ledger for personal and shared interests. 

From the nature of distributed networks, peers can join, leave, and disconnect the network as they please\cite{imtiaz2019churn} \cite{kim2018measuring}, i.e., users turning off their nodes to upgrade their version. 
However, this tends to decrease when discussing proof of stake systems \cite{cortes2021discovering}.
Validators have incentives for actively participating in consensus, although they can also be penalized if they don't do it \cite{cortes2023autopsy}.   
This leads PoS blockchains' networks to be more stable in general. Nonetheless, they still represent Byzantine fault tolerance systems\cite{castro1999practical}, where the system can overcome at some degree the sudden decrease of the honest participants ensuring a unanimous consensus over the state of the chain.
Beyond a fault-tolerant consensus mechanism, Ethereum adds a second layer of resilience by having multiple software to participate in the network. Each one is written in a different language and by a different team, targeting various end-users. 
Thus, they are optimized for different situations. 

Ethereum aims to be light and portable, substantially reducing the hardware requirements that the prior PoW meant. It has been proved \cite{cortes2021resource} that the new PoS version of Ethereum can be successfully run on medium-low hardware machines, with some clients needing less than 4GB of RAM and two cores to participate in the CL network successfully.
However, no prior work analyzes the impact on the performance of such a wide variety of end-users. The question of whether hardware-resource optimal clients, which allow solo stackers to validate from home, can perform as well as less hardware-restricted ones is still unanswered. 

Previous studies like \cite{kiffer2021under} have demonstrated that the node's location directly affects the networking performance and, inevitably, the performance of PoW-based application's nodes.
Message distribution is essential in PoS systems such as Ethereum. Although each slot gives a time window of 12 seconds to propose a block, commit the attestations, and aggregate them, receiving half a second sooner or later the block can directly impact the attestation of validators, as their attestation could shift from, the block is valid, to there wasn't a block at all.  
This means nodes in regions far from the core of the network could be disadvantaged. Placing a client in a poorer connected region can incentivize other validators to keep concentrating in the same geographical locations. Suppose a latency increase can put at risk the performance of a validator (and the economic stimulus attached to it). In that case, the short window of action that the protocol suggests would incentivize the centralization of clients in regions with the counterpart that it increases the exposition to censorship of the local authorities.

This paper will present the Ethereum CL performance comparison results based on network latency in different locations. In the study, we empirically analyze the stability of the clients under the real network behaviors of Ethereum's mainnet and the Goerli testnet. We analyze the performance of the different implementations by comparing the quality of the accomplished validator duties, comparing the score of the blocks generated by the other implementations. Furthermore, we reproduced the experiments in different geographical locations to explore the impact of the message propagation latencies on the ability to perform consensus duties.

\section{Methodology}
\label{sec:methodology}
As Ethereum keeps all the interaction with the chain available on the public blockchain, debugging the performance of validators from different locations remains, in most cases, accessible. However, processing and indexing the necessary information to determine the performance of a client or a validator in a human-readable way imply reconstructing, in some occasions, the chain's status in the past, and this is not that easy to reproduce. In this section of the paper, we will introduce the basic fundaments of Ethereum to understand the methodology we used to measure the performance of Ethereum nodes, i.e., slot time utilization, block generation, or attestation flags. Furthermore, we will also introduce the support software we built and used to generate the data we will later discuss in Section \ref{sec:evaluation}.

\subsection{Scoring Ethereum CL's duties}
\label{subsec:cl-duties}
Shifting Ethereum's consensus mechanism to PoS unarguably increased the complexity of the protocol. Since the merge, only active validators in the beacon chain can participate in the consensus, having to do it at least once every epoch. Validator's block proposals, block attestations, and sync committee votes are the duties that ensure that the blockchain keeps adding blocks under a consensus. Thus, the quality of these duties determines how well each validator contributed to the consensus, which ultimately defines the reward they get. 

The consensus layer of Ethereum is organized in epochs (see Figure \ref{fig:slot_time} for reference). Each epoch contains $32$ time windows of $12$ seconds called \emph{slots} where a single validator elected from the RANDAO Reveal \cite{randao-code} algorithm has the chance to aggregate a new block to the beacon chain. Since the rest of the existing active validators must reach a consensus over each proposed block, splitting the epoch in $32$ slots helps reduce the computational load of processing the duties of $750.000+$ (at the time of writing this paper) active validators. Thus, the whole list of validators is divided into the $32$ slots and then into a maximum of $64$ committees. This way, each added block serves as the main unit of time where new historical data is added to the beacon chain.

\subsubsection{Attestations}
Attestations or votes are the statements each validator must make to help finalize\footnote{Finalization is used to express when a block has been validated by more than 66\% of the network and for over two entire epochs. It represents the moment when the data inside the blocks of that epoch is no longer mutable.} beacon epochs. 
Therefore, each committee's resulting votes are aggregated before adding them to the following proposed blocks. This helps considerably reduce the block size by keeping track of the duties and saves time for future block proposers as they only have to listen to the latest aggregations of each slot. 
Inside the participation of each validator, the following three main flags determine the ``quality" of the attestation:
\begin{itemize}
    \item Source: hash of the justified checkpoint\footnote{Checkpoints in the CL represent the Beacon State root of the epoch's first slot, including the result of the state transition from the previous epoch.} at the moment the attested block was proposed.  
    \item Target: hash of the first block at the epoch.
    \item Beacon block root: hash of the attested block.
\end{itemize}
Each validator has $32$ slots to produce these attestations, leading to a second parameter that interferes with the ``quality" of the attestation, the inclusion delay. The inclusion delay refers to the number of slots it took for an attestation to get included in a block after the attested one. This means that the optimal performance for a validator is to produce a vote with the three flags correct and include it in the next block, meaning an inclusion delay of 1 slot.

\subsubsection{Sync committees}
Since the \emph{Altair Hard Fork}~\cite{altair-hardfork}, sync committees were added to help light clients validate blocks without fully downloading and processing the beacon chain. Each sync committee comprises 512 randomly selected validators who sign new block headers every slot and rotate every 256 epochs (8192 slots).

\subsubsection{Block proposals}
In every slot, a single active validator has the chance to generate and propose a beacon block. When that moment arrives, the validator adds the needed metadata of the block with as many aggregated attestations as possible. With an upper limit of 128 aggregations that can fit into a single beacon block, the CL reward that the proposer gets directly depends on the quantity and quality of the included attestations. From the reward that each non-previously included attestation flag generates, there is a separate percentage that gets saved for the proposer of the block that includes it. Thus, the more new attestations we add to a block, the greater the reward it generates.
The same happens with the sync committee rewards; the block proposer gets a percentage of the total reward that the included sync committee duties generate. For this reason, the block proposer is incentivized to include as many sync committee duties as possible.


\subsection{Slot time ranges}
\begin{figure}[!ht]
    \centering
    \includegraphics[width=0.95\linewidth]{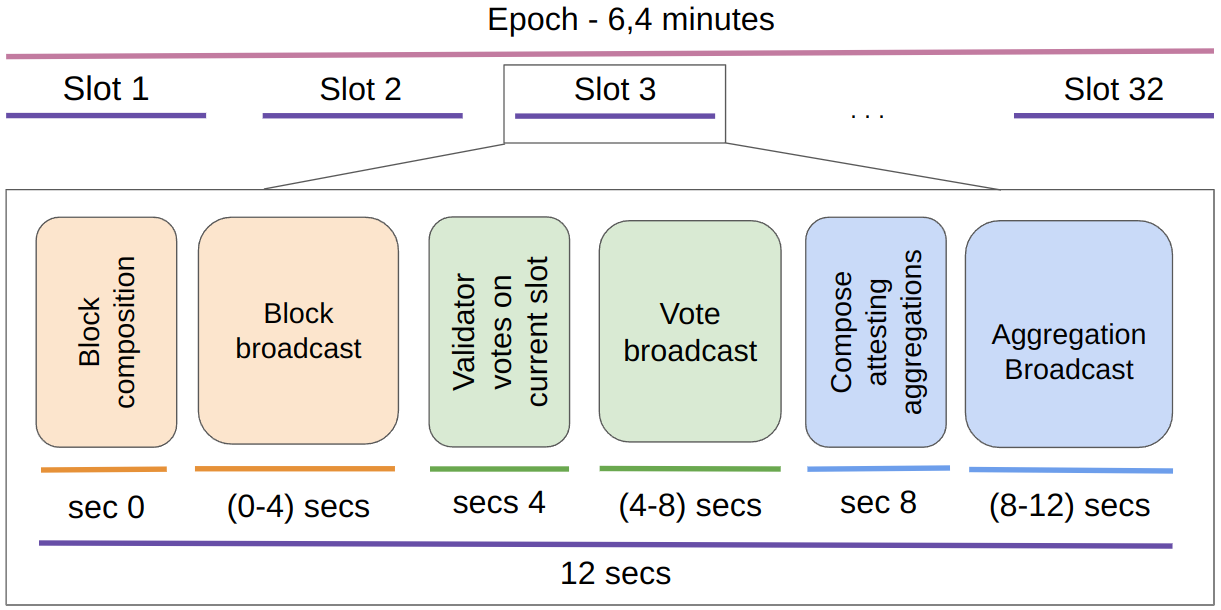}
    \caption{Slot time division between duties.}
    \label{fig:slot_time}
\end{figure}

We have already introduced the time division of Ethereum CL's blockchain. However, as Figure \ref{fig:slot_time} shows, there is still a smaller subdivision inside each slot. Although these numbers are just guidelines, following them is crucial to avoid generating confusion in the network. To achieve the best performance in the network, the following tasks need to be performed in order inside the slot:
\begin{enumerate}
    \item Block proposers are expected to create and broadcast a new block at the beginning of the slot (second 0 of the slot). This gives $4$ entire seconds for the message to reach the rest of the participants in the network. To do so, they have a time window of 4 seconds prior to the start of the slot to receive and group aggregated attestations from the previous $32$ slots. 
    \item After the first $4$ seconds of the slots, validators assigned to attest to it are expected to generate and broadcast their votes with their perception of the chain (attesting to the \emph{source}, \emph{target}, and \emph{head} they see). They share this vote with the corresponding beacon committee aggregators, and the spec assigns the same time range of $4$ seconds to broadcast the message.
    \item Finally, the committee aggregators must collect votes between seconds $4$ to $8$, producing the aggregated attestations at the $8$th second of the slot. In all committees, 16 validators are randomly selected to aggregate and broadcast the attestations. After that $8$th second, the network disposes of 4 extra seconds so that the next block proposer has enough time to receive all the aggregations.
\end{enumerate}
Keeping the correct timing between these tasks is crucial to avoid confusion in the network. For example, if a block proposer extends the creation of its block for $10$ seconds, the block could be received later than 12 seconds since the start of the slot, risking being voted as a missed block. If a validator waits too long to generate and send the attestation, the aggregators might not include that vote in the same slot, increasing the inclusion delay and reducing the final reward. 

\subsection{Support softwares}
\label{subsec:tools}
Although blockchains keep most of the interactions and balances publicly available on-chain, in some occasions, that information (i.e., validator duties) has to be reconstructed from the locally stored beacon states in the clients. As this information is essential to quantify and qualify the performance of a validator and client, we have relied on a set of tools that helped us gather and index all the necessary information. 

\subsubsection{Conensus rewards}

To compute the rewards obtained by a validator over the Maximum Extractable Reward (MER) on an epoch, we measured each validator's attestation and sync committee rewards on each epoch. We relied on the attestation and sync committee rewards models proposed by \cite{cortes2023autopsy} by using the same software \emph{GotEth} \cite{state-analyzer}, an open-sourced tool that indexes the following items from a trusted beacon node:
\begin{itemize}
    \item Validator individual duties.
    \item The quality of these duties (i.e., if validators missed a block proposal, the number of flags successfully voted).
    \item The max attestation and sync committee (if the validator during that epoch was inside a sync committee) reward that each validator could have achieved.
\end{itemize}

\subsubsection{Consensus block scorer}
\label{subsec:block-scorer}

To measure the capabilities of the different clients to generate beacon blocks, we created a custom open-sourced tool 
based on the beacon node multiplexer \emph{Vouch} \cite{vouch}. This custom tool can be connected to as many beacon nodes as we want, indexing some metrics from the live network into a PostgreSQL database. The tool can communicate with these provided beacon nodes, requesting them to generate a block at the beginning of every slot. With the final intention of analyzing the content of each proposed block, the tool aggregates the number of the included new votes, sync aggregates, attester slashing, and proposer slashing, generating a synthetic scoring system that later on will be used to compare them.
The score is derived directly from the beacon chain rewards formulas, removing the actual \emph{Base Reward} from the equation to make the score calculation faster. 
Furthermore, the tool can stream and record some events from each beacon node's API. For example, the tool tracks and timestamps every time it gets notified when a new block message is received. This allows us to compare the arrival time of messages such as new blocks.


\section{Evaluation}
\label{sec:evaluation}
To measure the performance of each Ethereum node, we deployed a set of experiments that would allow us to compare the results fairly. To keep the experiments away from simulations, we relied on Ethereum's live networks to perform these experiments. We chose Ethereum's mainnet as a mature, stable, and reliable network. However, since activating a validator in mainnet requires a deposit of $32$ ETH, we relied on Ethereum's Goerli testnet to activate $3000$ validators with the help of the EF~\cite{EF}. 

\begin{figure*}
    \minipage{0.32\textwidth}%
        \includegraphics[width=\linewidth]{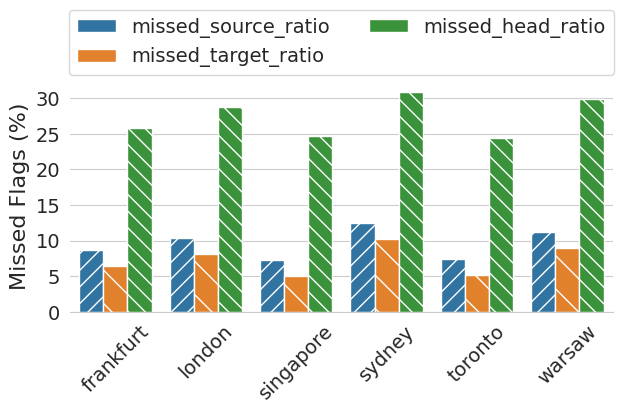}
        \caption{Average missed attestation flags.}
        \label{fig:avg-missed-flags}  
    \endminipage\hfill
    \minipage{0.32\textwidth}%
        \includegraphics[width=\linewidth]{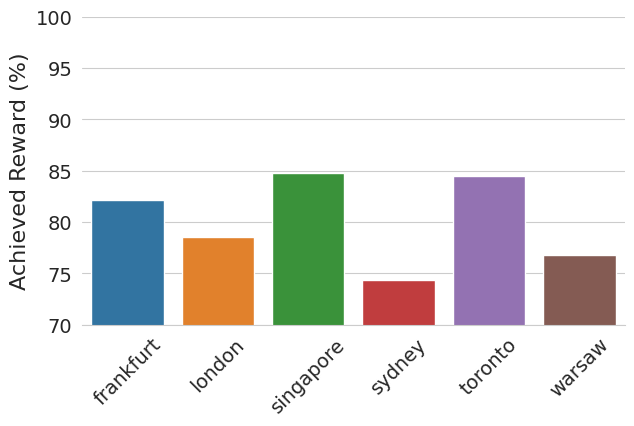}
        \caption{Average achieved reward per location.}
        \label{fig:avg-rewards} 
    \endminipage\hfill
    \minipage{0.32\textwidth}%
          \includegraphics[width=\linewidth]{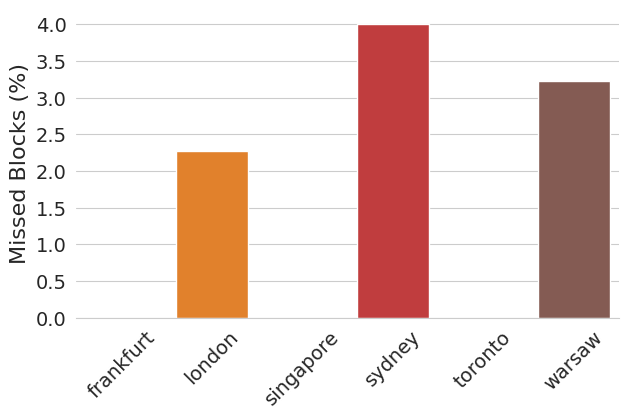}
        \caption{Aggregated missed blocks per location.}
        \label{fig:missed-blocks}
    \endminipage
\end{figure*}

The correctness of the attestation flags significantly impacts validators' rewards. To investigate why a single location would achieve fewer rewards than the rest, Figure \ref{fig:avg-missed-flags} shows the ratio of missed flags aggregated by location. 
We performed two main experiments divided into two sections \ref{subsec:validator-performance} and \ref{subsec:block-score-eval}. The first will evaluate the accomplishment of the validator's duties, and the second will introduce the differences when composing block proposals. 
Overall, for both studies, the control clients were grouped in groups of five. Each of the main available clients was paired with an EL client, which became mandatory after the merge. Thus, we spawned five pairs of CL clients + Nethermind \cite{nethermind} in four to six locations. However, each of the following sections will further introduce its configuration details.  

\subsection{Validator performance}
\label{subsec:validator-performance}
The current PoS consensus mechanism drastically changed the reward system in Ethereum. The protocol prioritizes rewarding validators' stability and continuous duty compliance. The reward retrieved through attestations represents the $61$\% of the gross reward a single validator can achieve. Thus, this first experiment compares the stability of performing attestations. To replicate the study and measure the impact of running clients in what we consider regions with more significant latency, Table \ref{tab:location_hardware_rewards} includes the configuration of nodes we designed to perform the study.  

Most cloud service providers cannot offer the exact same hardware resources in all the regions that we wanted to test. Therefore, we used multiple cloud providers with different capabilities. To ensure that all clients had roughly the same hardware resources, we broke the set of clients into two different machines in locations such as Sydney, Singapore, and Toronto. The hardware limitations outside the EU and the US were noticeable. 

\begin{table}[]
    \centering
    \caption{Hardware setup per location for the rewards study.}
    \begin{tabular}{cccccc}
        \hline
        City & Hardware & CPU & Memory & Storage & Clients \\
        \hline  
        Frankfurt & Baremetal & 16c. & 64GB & 1.9TB NVMe & All \\
        London & Baremetal & 16c. & 64GB & 1.9TB NVMe & All \\
        Warsaw & Baremetal & 16c. & 64GB & 1.9TB NVMe & All \\
        Sydney1 & VM & 16c. & 32GB & 850GB SSD & LH,L,P\\
        Sydney2 & VM & 8c. & 16GB & 700GB SSD & N,T  \\
        Singapore1 & VM & 16c. & 16GB & 900GB SSD &  LH,L,P \\
        Singapore2 & VM & 4c. & 16GB & 700GB SSD & N,T \\
        Toronto1 & VM & 16c. & 32GB & 910GB SSD & LH,L,P \\
        Toronto2 & VM & 4c. & 16GB & 660GB SSD  & N,T \\
        \hline
    \end{tabular}
    \label{tab:location_hardware_rewards}
\end{table}
\begin{table}[]
    \centering
    \caption{Table with the versions used during the studies.}
    \begin{tabular}{ccc}
        \hline
        Short & Client & Version \\
        \hline  
        LH & Lighthouse & v3.1.2-01e84b7 \\  
        L & Lodestar & v1.2.1/5813d39 \\  
        N & Nimbus & v22.9.1-ad7541-stateofus \\  
        P & Prysm & v3.1.1 \\  
        T & Teku & 22.9.1 \\
        \hline
    \end{tabular}
    \label{tab:client_hardware_version}
\end{table}

\subsubsection{Reward comparison based on Goerli validators }

The first part of the study compares the aggregated validator rewards per location between epochs $157835$ and $158835$, or in a human-readable format, between dates February 23rd 2023, and February 27th 2023. Intending to discover any hints of a possible underperformance of any specific region, Figure \ref{fig:avg-rewards} shows the achieved reward by the aggregated validators per location out of their respective MER. 

Considering the aggregation of the validators per location as our validator control pools, we observe that most nodes achieved a similar reward when comparing it with their MER. It is expected to see drops in the achieved rewards when we compare validators in testnets with validators in mainnet. In this case, the Goerli testnet is publicly open for participants to collaborate without needing fiat collateral to ensure their participation. Thus, there is a higher ratio of participating nodes that are not properly maintained, wrongly configured, or directly disconnected. This ultimately impacts the stability of the network, experiencing more missed blocks, more reorganizations, more missed flags, and thus bigger inclusion delays lowering the MER compared with validators in mainnet. 

With all these said, in Figure \ref{fig:avg-rewards}, we still find a slight variation around an average reward of $80.2$\% for most locations. The most significant exceptions are recorded in Sydney and Warsaw, which fall to $74.3$\% and $76.8$\%, respectively. 
As explained before, the instance deployment differs in some locations. 
Nodes located in Frankfurt, London, and Warsaw share a similar infrastructure setup, in which the achieved reward is similar, varying between $76$\% and $82$\%.
Nodes located in Singapore, Sydney, and Toronto also share a similar infrastructure setup between each other. However, Sydney achieved $10$\% less MER than the other two locations ($84$\%). 
Since Sydney is the most remote location of the chosen setup, as most nodes concentrate in Europe and North America \cite{cortes2021discovering}, a message is expected to have a slightly higher latency to reach nodes in such remote locations. Thus, this could show that network latency significantly impacts performance. It is remarkable that validators hosted in nodes with apparently ``worst" connections, i.e., nodes in Singapore, extract more rewards than ``better" connected ones, such as nodes in Frankfurt. This indicates that hardware also plays a vital role in achieving a more significant share of the achievable rewards.  

\paragraph{Missed heads}
The head flag inside the attestation points to the head slot in the canonical chain. To send a correct head attestation, the validator must point to the head root and provide this attestation with an inclusion delay of 1 block. Otherwise, the head attestation is given as wrong. This explains why it is the most commonly failed flag among validators in the network.
Failing the head attestation flag could mean that the node falls more into reorgs or that the attestation is not sent in time and, thus, not included in the next block (inclusion delay=1). Figure \ref{fig:avg-missed-flags} shows the average flag failure per location, where we can see that the average head attestation flags' failure rate is $27.4$\%. The figure shows that nodes in London, Sydney, and Warsaw failed over that average $28.7$\%, $29.9$\%, and $30.8$\% missed head flags, respectively.

\begin{figure*}
    \minipage{0.30\textwidth}%
    \includegraphics[width=\linewidth]{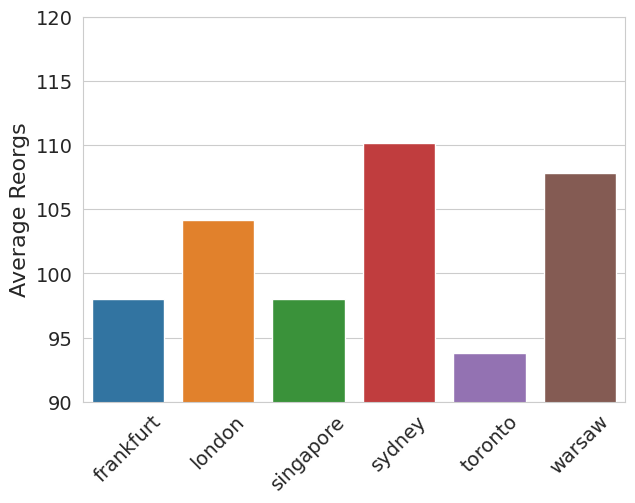}
    \caption{Number of chain reorganizations per location.}
    \label{fig:goerli-reorgs}
    \endminipage\hfill
    \minipage{0.33\textwidth}%
        \includegraphics[width=\linewidth]{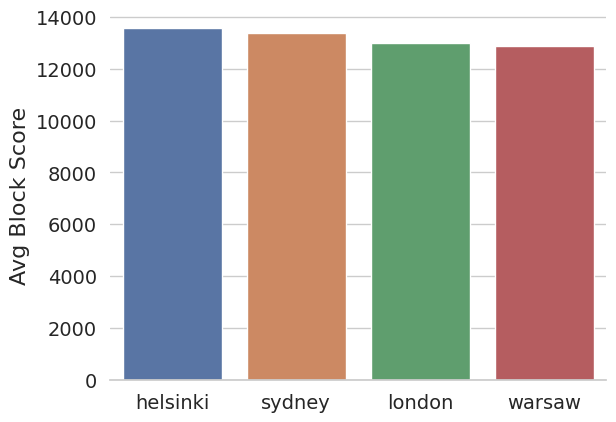}
        \caption{Average block score per location.}
        \label{fig:average-score-location}  
    \endminipage\hfill
    \minipage{0.33\textwidth}%
        \includegraphics[width=\linewidth]{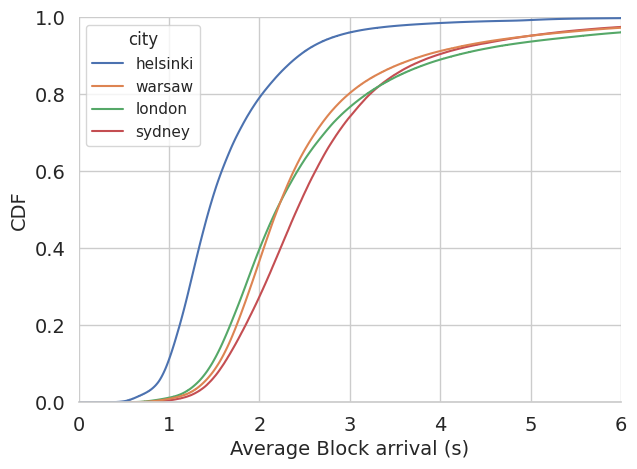}
        \caption{Average arrival latency of block messages inside slot time per location.}
        \label{fig:arrival-time-per-location}  
    \endminipage
\end{figure*}

\paragraph{Missed targets}
Conversely, the target attestation flag is the least failed flag and the one that brings the most rewards out of all three flags. This is why we define that if the target flag is failed, the node is most likely out of sync and has been like this for a while.
Figure \ref{fig:avg-missed-flags} shows a similar pattern for the target flags, with nodes in Frankfurt, Toronto, and Singapore missing them $5$\% and $6.3$\% of the time. On the other hand, London, Warsaw, and Sydney nodes stay beyond that average, reaching the missed ratio of $8$\% $8.9$\% and $10.2$\%, respectively. 
It is clear, then, by analyzing the aggregation between the number of missed head and target flags, that despite Sydney sharing the same hardware with Toronto and Singapore, it falls out of sync almost double the times. 

\subsubsection{Proposer duties}
Validators also earn rewards from proposing blocks when they are randomly chosen. Block rewards are sporadic but very high, so it is frustrating for validator owners to miss the chance of gaining such a high reward with a straightforward duty.
When proposing a new block, the node must be fully synced with the network and follow the chain head without delays. Not doing so could cause the validator to miss the block proposal (or do it very late), missing out the substantial block rewards it generates.

As block proposers are randomly chosen at every epoch, Figure \ref{fig:missed-blocks} shows the aggregated ratio of missed proposer duties between locations.
The figure shows that pool nodes on each location failed an average of $1,66$\% of block proposal duties, with Sydney nodes failing up to $4$\% of the block proposal duties, while Frankfurt, Toronto, and Singapore didn't miss any of the proposals.
Once again, it is most likely that Sydney nodes tend to fall more into the out-of-sync state and, therefore, can not perform their duties in time.

\paragraph{Chain reorganizations on clients}
Nodes have different behaviors depending on the hardware they are running on and the location where they are placed on. However, validators' achieved rewards or the ratio of missed duties are not the only methods to measure the performance of a node. The high ratio of missed target flags when attesting is the first indicator of a stability problem on a particular node, or in this case, in a location. 
It is hard for a validator to perform its duties correctly when the underneath node is not fully synced with the chain. Thus, we can interpret that nodes in Sydney, as a representation of less well-connected nodes, tend to lose synchronization more often. We have chosen reorgs (chain reorganizations) as a way of measuring the stability of a node. A chain reorg represents having to drop a number of blocks (with their states) and sync the canonical version of them because the node was in a non-valid variation of the chain. As the late arrival of messages can generate those minor forks, Figure \ref{fig:goerli-reorgs} shows the aggregation of reorgs registered in each location.
In the graph, we can observe that Sydney has the biggest reorg average from the six different places, with eight more registered events than the average of $102$ of all the locations. Without being an extremely large number, we can attribute the differences between duties accomplishment to the fact that the reorgs can also be defined by the number of blocks you had to drop and resync. Unfortunately, the node does not offer this information to us. We will discuss this step in detail in the following section.  





\subsection{Block scores}
\label{subsec:block-score-eval}

With the major differences we spotted between the arrival times of the different instances deployed in \emph{mainnet} (the most stable network in the Ethereum ecosystem), we wanted to study the impact of this arrival latency on the capabilities of the clients to compose blocks. Following in order the rewards quantities that a validator received over time in the Consensus Layer, the resulting rewards of block proposals follow attestations with a $7,6$\% of the total reward. Thus, we decided to benchmark, with a synthetic block score (\ref{subsec:block-scorer}), the differences that each client and each location produce. 
We experienced similar problems to the ones deploying the validator rewards study; fitting five CL clients with their respective five Nethermind clients under the same or similar machines was unmanageable. Thus, the clients and the machines were organized and deployed as Table \ref{tab:location_hardware_scores} summarizes. The data displayed and analyzed in the following sections belongs to the range of slots $5760722$ to $5888722$, which belongs to the range from February 9th, 2023, to February 27th, 2023.

\begin{table}[]
    \centering
    \caption{Hardware type on each tested location of the block score study.}
    \begin{tabular}{cccccc}
        \hline
        City & Hardware & CPU & Memory & Disk & Clients \\
        \hline  
        Helsinki & Baremetal & 32c. & 128GB & 10TB NVMe & All \\
        London & VM & 32c. & 128GB & 7TB SSD & All \\
        Warsaw & VM & 32c. & 128GB & 7TB SSD & All \\
        Sydney1 & VM & 4c. & 32GB & 2.5TB NVMe & LH,L,P \\
        Sydney2 & VM & 4c. & 32GB & 1TB NVMe & N \\
        Sydney3 & VM & 4c. & 32GB & 1TB NVMe & T \\
        \hline
    \end{tabular}
    \label{tab:location_hardware_scores}
\end{table}

\subsubsection{Beacon block generation differences}
\label{subsec:beacon-block-generation}
Figure \ref{fig:average-score-location} shows the aggregated score of each block across the clients of each location, where we appreciate insignificant differences. Once again, there is a clear dominance of nodes located in Helsinki, outstanding over nodes in Sydney, London, and Warsaw that achieved $1.37$\%, $4.29$\%, and $5.16$\% less score, respectively.
Although each of the locations should have enough time to get the same attestation aggregations in the course of the last $4$ seconds of a slot, different factors could produce this difference in the average score of the blocks:

\paragraph{Message propagation latency} 
Higher latencies when receiving messages clearly limit the number of aggregations that you can include in a block. Figure \ref{fig:arrival-time-per-location} displays the Cumulative Distribution Function (CDF) of the message arrival time in each location, where the Y axis represents the normalized percentiles between ranges $0$ and $1$, and the X axis the arrival time in seconds. We can read the figure as the $50$th percentile of blocks ($0.5$ on the Y axis) in Helsinki arrived in $1.44$ seconds or less. We can see that, despite London having the second-best median of arrival times, $2.18$ seconds, it has one of the worst $90$th percentiles of $4.15$ seconds. The large tail of block arrivals beyond $4$ seconds represents $10$\% of the total tracked messages and it partially explains the block score differences between locations.      
\begin{figure*}
    \minipage{0.32\textwidth}%
        \includegraphics[width=\linewidth]{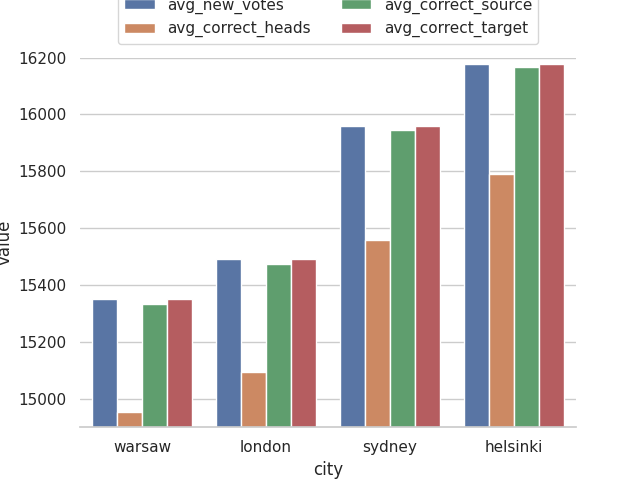}
        \caption{Number of new votes and their correct flags on each location.}
        \label{fig:new-votes}
    \endminipage\hfill
    \minipage{0.33\textwidth}%
        \includegraphics[width=\linewidth]{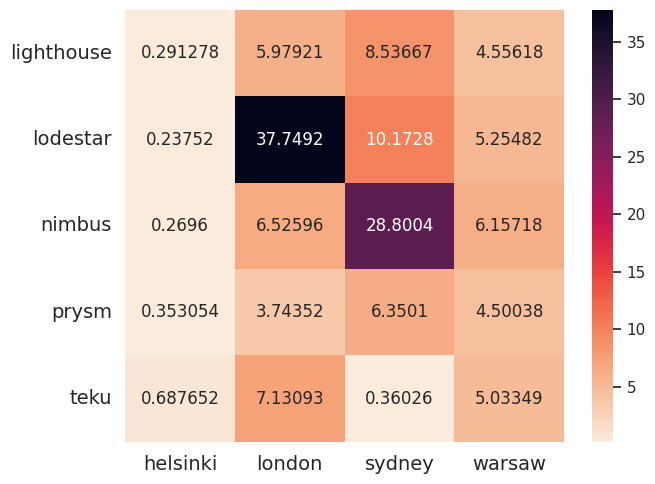}
        \caption{Percentage of slots each beacon node was out of sync in each location.}
        \label{fig:out-of-sync}
    \endminipage\hfill
    \minipage{0.34\textwidth}%
        \includegraphics[width=\linewidth]{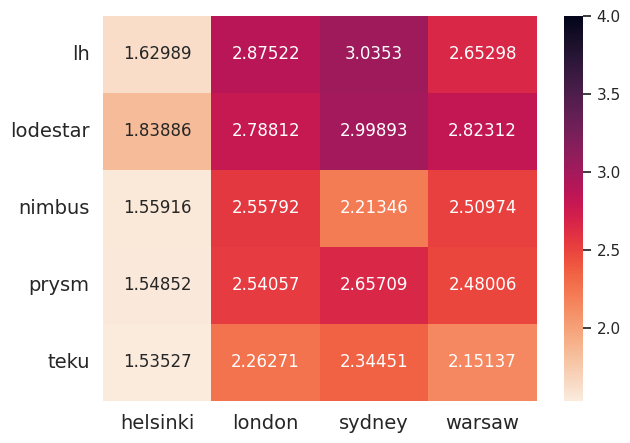}
        \caption{Average block arrival time per client and location.}
        \label{fig:block-arrival-latencies}
    \endminipage
\end{figure*}

\paragraph{The aggregation of more new votes in a block} 
Receiving messages later means adding fewer votes in a new block, which are the ones producing the rewards for the block proposer. Figure \ref{fig:new-votes} displays the average number of new votes included at each location, including the average of their correctness. The figure clearly shows the dominance of nodes in Helsinki that not only aggregate more new votes but also have, on average, more correct attestation flags.   

\paragraph{Desynchronization of beacon nodes} 
Desynchronization is, even if we put endless effort into avoiding it, one of the major drawbacks we found to explain the difference in average score between locations. Of course, a beacon node can not process new messages and generate new blocks if it gets out of sync with the head of the chain. Several events could cause this to happen, such as big local reorgs caused by higher latencies or by hardware limitations (slow disk where to prune states or not enough CPU available to validate messages on time). Figure \ref{fig:out-of-sync} displays the percentage of slots each node was down versus the number of slots we measured. In the picture, we can appreciate that Helsinki was barely unsynchronized. We can also appreciate how Lodestar in London affected the previous averages, with a $37,74$\% of the measured slots out of sync. The results are more puzzling when we compare it with Warsaw, concluding that despite having similar hardware, it was stable and in harmony with the rest of the clients at around $5,25$\% downtime, and despite this, it performs worse than London both in correct flags and block score.

\subsubsection{Latencies on block arrival}
\label{subsubsec:block-arrival-latencies}

We have already introduced the importance of message broadcasting latencies within the slot time range. Shorter notice of messages such as blocks or aggregations might be critical to ensure that validator duties are correctly achieved. As we would imagine, different locations with different connections with the rest of the network could generate different network perspectives. This way, if the defined 4 seconds to distribute a message weren’t enough, we could expect that those regions with higher latencies would perform poorer than the better-connected ones. 
To corroborate our hypothesis, we tracked the arrival of block messages to each client in the four different mainnet instances. The tool would first subscribe to the beacon node’s API to stream the arrival of new blocks, indexing the event with the notification timestamp. From the difference between the local timestamp and the time the slot started (second ``0” of the slot, when block proposers should publish the block), we have aggregated the arrival times, distinguishing the distribution from the following figure.
Figure \ref{fig:block-arrival-latencies} shows how clients in Helsinki received the block messages significantly sooner. We would expect from previous experiences that the European area has better connectivity in general terms, and this graph proves it: on average, nodes in Helsinki received blocks $1.06$ seconds sooner than nodes in Sydney. However, even though we are not making any distinction across clients (all clients were aggregated) to have a fair comparison between the locations, the differences between London, Sydney, and Warsaw are not as remarkable as the arrival times in Helsinki. With an average block arrival time of $2.60$, $2.68$, and $2.53$ for each, the figure shows that the major differences between the instances could be originated from the distinct machines chosen per location. 

Although we tried to have the most similar machines in each location, Cloud Service providers couldn’t offer the same hardware tier across the four locations. With a clearer dominance of resources from the machine in Helsinki, even though it had to share the disk among ten different clients (five CL + five EL), it still had a powerful bare metal machine with $32$ CPU cores and $128$GB of memory. Limited CPU resources could become a bottleneck when a client receives and validates a block message, as they generate computational spikes every $4$ seconds. Thus, to some degree, this CPU bottleneck can increase the measured latency of processing blocks, including the time our tool got notified. 
Of course, different clients mean different implementations. The heat map in Figure \ref{fig:block-arrival-latencies} breaks down the block arrival times per client and location showing some differences across the clients. Without being highly dispersed among the clients, it is clear that some clients received and processed the block faster than others during our study. Despite showing the same distribution as the previous figure \ref{fig:out-of-sync} among the different locations, Lodestar is in the order of $300$ms slower in receiving and processing new blocks. 

There are many reasons that could cause a later message arrival. On the one hand, we have the number of connected peers, where Lodestar and Prysm stay at $50$ to $55$ number of simultaneously connected peers, Lighthouse and Teku stay with an average of $80$ to $100$ connections and Nimbus outstands the rest with $160$ direct peers. However, on the other hand, clients must allocate more computational power at the arrival of messages to process and validate the messages or to update the local beacon state. For this reason, the bigger the number of simultaneous connections the node has, the more messages you need to process. Thus, this normally generates CPU usage peaks every $4$ seconds, and as it happens with higher latencies, both have a direct impact on the performance of the client.

\section{Discussion}
\label{sec:discussion}
This paper presents a distinct methodology that can be used to analyze the impact of latency on the performance of Ethereum CL nodes located around different geographical locations. 
In the presented results, we have identified that Ethereum CL's default $4$ seconds for broadcasting gives enough margin to propagate and receive all the necessary messages (i.e., sync committee attestations and block proposals) in most geographical locations, at least if the clients are running in instances with minimum hardware specifications.

We have empirically demonstrated that some regions, such as Oceanía and Southeast Asia, have higher latency distributions when receiving block messages for the first time; with some locations receiving 10\% of the messages beyond the $4$ second mark. This makes nodes more likely to lose the correct head of the chain, leading more often to reorg their local chain and ultimately failing or performing duties more poorly because the node is not fully operational. This gets mirrored when comparing the rewards achieved between the available locations, where Sydney nodes earned $10$\% fewer rewards from the MER than the rest. 
We have demonstrated that even though the hardware requirements to participate in a Post-Merge Ethereum are substantially smaller than its predecessor PoW consensus, there are some minimum requirements to run both EL and CL clients without a hardware bottleneck.   

Furthermore, the paper compares the networking performance of the different available Ethereum CL clients. We have demonstrated that under optimal networking and hardware conditions (i.e., the instance in Helsinki), there are barely any differences between clients, i.e., a similar downtime or out-of-sync time.
We have identified that hardware limitations directly increase the latency (mostly from message validation). Thus, nodes face more downtime as they might get out of sync, or could potentially add fewer new votes to their blocks.

\section{Conclusions and Future work }
\label{sec:conclusion}
This paper presents a new methodology to quantify, qualify, and compare the performance of Ethereum beacon nodes. We have demonstrated that despite all the clients performing similarly under optimal hardware and networking conditions, variations or limitations on these same ones can severely impact the stability of the beacon nodes, and thus, the performance of the hosted validators. 
We can conclude the study by stating that there is indeed a performance impact related to the connectivity of a node. Still, it is only significant when the hardware is not properly dimensioned. With the presented clear difference in block arrival latencies across locations, nodes further away from the core of the network can see their stability and performance reduced. Reaching even critical stability problems if the hardware is not slightly over-dimensioned. 
In future work, we aim to explore a real-time model able to aggregate and compare all the presented parameters to monitor and alert the performance of a client with or without a validator.

\section{Acknowledgements}
\label{sec:ack}

This work has been supported by the Lido Ecosystem Grant Organization (LEGO), the Ethereum Foundation under the Research Grant FY21-0356, and Protocol Labs under its Ph.D. Fellowship Program FY22-P2P. We want to thank the researchers from Attestant.io for helping with the necessary infrastructure and discussions. Also, to Paristosh from the Ethereum Foundation, for his implication and help in activating the Goerli validators. In particular, Izzy and Alvaro Revuelta for their constructive feedback on this study.

\bibliographystyle{IEEEtran}
\bibliography{references}

\end{document}